\documentclass[pdftex, a4paper]{scrartcl}
\usepackage[applemac]{inputenc}
\usepackage[T1]{fontenc}
\usepackage{amsmath}
\usepackage{color}
\usepackage{braket}
\usepackage{amsfonts}
\usepackage{graphicx}
\usepackage{hyperref}
\hypersetup{colorlinks= true, allcolors=blue}

\title{A Decoherence--Based Approach to the Classical Limit in Bohm's Theory}
\author{Davide Romano\thanks{Centre of Philosophy, School of Arts and Humanities, University of Lisbon. Address: Alameda da Universidade, 1600-214, Lisbon, Portugal. E-mail: davide.romano@edu.ulisboa.pt}}

\begin{document}
\maketitle

\begin{center}
Article published in the special issue edited by A. Drezet: \emph{Pilot-wave and beyond: Louis de Broglie and David Bohm's quest for a quantum ontology}, Foundations of Physics.  
\end{center}

\begin{abstract}
The paper explains why the de Broglie-Bohm theory reduces to Newtonian mechanics in the macroscopic classical limit. The quantum-to-classical transition is based on three steps: (i) interaction with the environment produces effectively factorized states, leading to the formation of \emph{effective wave functions} and hence \emph{decoherence}; (ii) the effective wave functions selected by the environment--the pointer states of decoherence theory--will be well-localized wave packets, typically Gaussian states; (iii) the quantum potential of a Gaussian state becomes negligible under standard classicality conditions; therefore, the effective wave function will move according to Newtonian mechanics in the correct classical limit.  As a result, a Bohmian system in interaction with the environment will be described by an effective Gaussian state and--when the system is macroscopic--it will move according to Newtonian mechanics.

\end{abstract}
\clearpage

\tableofcontents
\vspace{10mm}

\section{The strategy of this paper}
This paper proposes a solution to the classical limit problem, i.e. the problem of recovering Newtonian mechanics from quantum mechanics in the macroscopic regime. In this regime indeed we expect ordinary classical objects such as tables, chairs and ourselves to emerge from the fundamental quantum description. \\
The starting point of this work is the consideration that the classical limit is necessarily characterized by decoherence\footnote{The importance of decoherence for the classical limit can hardly be ignored nowadays. Standard references are, for example: Joos et al. (2013), Schlosshauer (2007) and Zurek (2002).} as any quantum system (especially macroscopic systems), in normal conditions,\footnote{As long as they are isolated, even macroscopic systems maintain their quantum behavior, such as the VIRGO interferometer for the detection of gravitational waves.} interacts spontaneously with its surrounding environment. Think for example of a table, a chair or a cat that scatters off the surrounding air molecules and interacts with light photons, cosmic rays, cosmic microwave background radiation.
Also, given the rapidity for decoherence to become effective in the macroscopic limit, \footnote{Typical decoherence times for a dust grain in collisional decoherence range from $10^{-18}-10^{-31}s$. See, for example, Schlosshauer (2007, sect. 3.4, p. 135)} the interaction with the environment must be taken into account as a primary factor for the transition from quantum mechanics to classical mechanics. \\

\noindent The paper presents a decoherence-based strategy for deriving Newtonian mechanics from quantum mechanics, using Bohm's theory as a framework for quantum mechanics.\footnote{I have analyzed the limits of decoherence theory in the standard context in Romano (2022). } One remark is in order: it is argued sometimes that in Bohm's theory decoherence may not play a central role in the quantum to classical transition, given the common particle ontology between Bohm's theory and Newton's theory. However, this is not correct. The motion of the Bohmian particles is indeed completely determined by the wave function. The only way to change the Bohmian trajectories and make them look Newtonian in the classical regime is by a modification of the wave function that guides those trajectories. In other words: the quantum to classical transition in Bohm's theory is first and foremost a transition about the wave function.\footnote{This consideration is independent from the specific interpretation of the wave function that we adopt in Bohm's theory. Different interpretations will simply draw the line between the formalism and the ontology at different points of the proposed strategy.} This transition is realized by some physical process that produces a fundamental modification of the wave function, which in turn produces a modification of the Bohmian particle's behavior. It is the aim of this paper to individuate the physical process that realizes this transition and to explain why the Bohmian trajectories do become (approximately) Newtonian in the macroscopic limit.  \\ 
The strategy of the paper is based on the following three steps:

\begin{description}

\item [1. Interaction with the environment.]  We consider a macroscopic system that interacts with external degrees of freedom, collectively called ``the environment''. This interaction produces a system-environment entangled state. When the environment is composed of a very large number of particles,\footnote{In the macroscopic regime, the number of particles composing the system or the environment are typically characterized by the Avogadro's number: $10^{23}$.} the entangled state will be effectively factorized, leading to the formation of system's and environment's effective wave functions and thus decoherence. Decoherence, i.e. the impossibility to detect quantum interference between different components of the (sub)system over appreciably distant regions of space, is the result of the formation of the effective wave functions. In other words: it is not decoherence that produces effective wave functions but the other way round: it is the formation of effective wave functions that produces decoherence. The latter is an empirical result, the former is its physical explanation. 

\item [2. Gaussian states] The effective wave functions produced by the interaction with the environment, which we may call \emph{environmentally-selected effective wave functions} (ES-EWFs) are well-localized states. This follows from the diagonalization of the reduced density matrix described by the subsystem master equations\footnote{A master equation is the dynamical equation that describes the evolution of the subsystem (represented by a reduced density matrix) in interaction with the environment.}, which transforms an initial pure state into an (improper) mixture of well-localized states.\footnote{For the description of diagonalization process, see Schlosshauer (2019, pp. 7-8). For a critical survey of the interpretation of this process in the standard context, see Romano (2022, sect. 2).} Such states are generally called \emph{pointer states}: these are the states that survive the decoherence process and that remain stable during the interaction with the environment. We thus recognize that the effective wave functions selected by the environment (ES-EWFs) in Bohm's theory correspond to the pointer states selected by the environment in decoherence theory. In particular, in the macroscopic limit we have strong indications that the pointer states, and thus the ES-EWFs, will be Gaussian states. This result follows from the analysis by Zurek \emph{et al.} (1993) of the Quantum Brownian Motion (QBM) and will be discussed in section 5.

\item [3. Newtonian dynamics] From steps (1) and (2), it follows that a Bohmian system in interaction with the environment is described by an ES-EWF, mathematically represented by a Gaussian state. This system will move according to Newtonian mechanics when the standard macroscopic conditions ($m \rightarrow \infty$, $\hbar \rightarrow 0$) apply. Under these conditions indeed the quantum potential \footnote{The quantum potential (and the quantum force) may be seen as a novel physical potential (and force) or just as part of the mathematical description for the dynamics of the Bohmian particles. My position on this topic has been developed in Romano (2021), but the strategy presented in this paper is independent from the specific interpretation of the quantum potential (and more generally, of the wave function) that one takes in Bohm's theory.} of the ES-EWF is approximately zero and, consequently, the Bohmian dynamics reduces (approximately) to the Newtonian dynamics.\footnote{Under these conditions, the Bohmian trajectories will be approximately and not exactly Newtonian trajectories, but with a level of approximation that is impossible to detect in the macroscopic regime.}  That is: a macroscopic Bohmian system in interaction with the environment will move according to classical Newtonian mechanics

\end{description}

\section{Bohm's theory}
In this section, I briefly review the formalism of Bohm's theory, a version of quantum mechanics where systems are described by the wave function and  point-particles, the latter being represented by their position in three-dimensional space.

\noindent Bohm's theory (Bohm (1952a); Bohm \& Hiley (1993)), also called pilot-wave theory or de Broglie--Bohm theory, is a quantum theory of particles in motion: the state of an N-particle Bohmian system is represented by the N-particle wave function $\psi (x_1, x_2, \ldots, x_N)$, solution of the Schr\"odinger equation, and the particle configuration $\mathcal{Q}=(q_1, q_2, ..., q_N)$, where $q_i$ represents the actual position of the $i$-th particle. The dynamical equations of Bohm's theory can be derived by the following procedure. First, we decompose the wave function in polar form: 
\begin{equation}
 \psi(x,t)=R(x,t)e^{\frac{i}{\hbar}{S(x,t)}}
 \label{polar}
 \end{equation}
then, inserting \eqref{polar} into the Schr\"odinger equation, and separating the real and imaginary part, we obtain two coupled dynamical equations for the real fields $R(x,t)$ and $S(x,t)$, respectively the amplitude and the phase of the wave function: 
\begin{equation}
\frac{\partial{S}}{\partial{t}}+\frac{(\nabla{S})^2}{2m}+V+Q=0
\label{qHamilton}
\end{equation}

with 

\begin{equation}
Q=-\frac{\hbar^2}{2m}\frac{\nabla^2{R}}{R} 
\label{qpotential}
\end{equation}

and

\begin{equation}
\frac{\partial{R^2}}{\partial{t}}+\nabla \cdot \left(R^{2}\frac{\nabla{S}}{m}\right)=0 
\label{continuity}
\end{equation}

\noindent Eq.\eqref{qHamilton} is called the quantum Hamilton--Jacobi equation, since it has the same structure of the classical Hamilton--Jacobi equation, apart from the ``quantum potential'' term $Q$. blueAs for the classical case, the quantum Hamilton-Jacobi equation defines a real scalar field $S(x,t)$ on configuration space and the particles' trajectories are the integral curves of $S(x,t)$. The velocity of the particles can be derived by the kinetic energy term: 

\begin{equation}
\frac{P^2}{2m}\equiv \frac{(\nabla{S})^2}{2m} \label{kinetic}
\end{equation}

\noindent From eq. \eqref{kinetic}, we see that the the particles' momentum is $P=\nabla{S}$ and thus the particles' velocity is defined by the following equation:

\begin{equation}
v=\frac{\nabla{S}}{m}
\label{guidingdue}
\end{equation}

\noindent which is called the \emph{guiding equation}, since the wave function ``guides'' the particles' motion. The term \eqref{qpotential} is called the \emph{quantum potential}, as it acts in the quantum Hamilton-Jacobi equation as an extra-potential term generated by the wave function through the amplitude $R(x,t).$\footnote{Note however that the amplitude $R(x,t)$ and phase $S(x,t)$ of the wave function are not independent terms, they are coupled together through the continuity equation \eqref{continuity}.} For an N-particle system, eq. \eqref{qHamilton} describes the motion of N particles with kinetic energy $E_K=\frac{(\nabla{S})^2}{2m}$ and affected by the classical and quantum potential ($V+Q$). The total energy of the system $E_{tot}$ is thus given by the sum of three terms: the kinetic energy, the classical potential energy and quantum potential energy.\footnote{In the 1953 Einstein's example of the particle in a box, the Bohmian particle is at rest even if the system has a finite total energy but both the kinetic energy and the classical potential energy are zero inside the box. The energy is indeed absorbed by the quantum potential: $E_{tot}=Q$. See, on this point, Bohm (1952b, p. 184)}. I will analyze this example in the next section. 

\begin{equation}
E_{tot}= E_K + V + Q
\end{equation}

The dynamical equation \eqref{continuity} represents the continuity equation for $R^{2}=|\Psi|^2$, which describes the conservation of the Born's probability distribution $|\psi|^2$ through time. In Bohm's theory, the quantum probability distribution refers to the statistical distribution of the actual particles' positions. Given a system with wave function $\psi$, the particles' positions are statistically distributed according to $|\psi|^2$. That is: the Born's probability distribution reflects the epistemic ignorance about the actual particles' configuration. The continuity equation guarantees that, if the actual particles' positions are $|\psi|^2$-distributed at the initial time $t_0=0$, then this distribution will be preserved by the dynamics and so will be maintained at all later times $t>t_0$ .\footnote{Two different approaches have been proposed in the literature to explain why the initial particle configuration of a Bohmian system is distributed according to $|\psi|^2$: the typicality approach by D\"urr, Goldstein \& Zangh\`{i} (1992) and the relaxation dynamical approach by Valentini (1991). A comparative review of the two approaches has been made by Norsen (2018). See also Drezet (2021), for a recent proposal to justify the Born's rule using a decoherence framework.} \\Taking the gradient of both sides of eq.\eqref{guidingdue} and using eq.\eqref{qHamilton}, we obtain the particle acceleration:

\begin{equation}
m\ddot{x}= -\nabla{V}-\nabla{Q} 
\label{quantumnewtonuno}
\end{equation}

Defining, in analogy with Newton's theory, the quantum force as (minus) the gradient of the quantum potential: $F_Q=-\nabla{Q}$, we finally obtain the fundamental equation of Bohm's theory, i.e. the \emph{quantum Newton's law}:
\begin{equation}
m\ddot{x}=F_C+F_Q  
\label{quantumnewtondue}
\end{equation}

which describes the acceleration of a Bohmian particle, or a configuration of particles, generated by the sum of the classical and the quantum force.\footnote{For example: the quantum force is what makes the particles' trajectories deviate from straight lines in the two-slit experiment, even if there is no classical force acting on the Bohmian particles between the slits and the final screen.} Bohm's theory, as Newton's theory, is a second-order theory of particles in motion. And, exactly as in Newton's theory, the particles' motion is generated by physical potentials, or by forces generated by those potentials. From eq. \eqref{quantumnewtondue}, we see that Bohm's theory is indeed a generalization of Newton's theory: while, in the latter, particles are influenced only by classical potentials, in the former they are influenced by classical and quantum potentials. The quantum to classical transition is thus realized when the quantum potential and the quantum force are negligible. Under these conditions, indeed, the quantum Newton's law is approximately equivalent to Newton's second law of dynamics and the Bohmian particles will (approximately) move according to Newtonian mechanics. 

\begin{equation}
m\ddot{x}=F_C+F_Q \approx F_C \quad \textrm{when} \quad \left( Q \rightarrow 0, \nabla{Q} \rightarrow 0 \right)
\end{equation}

\section{Negligible quantum action: $\hbar\rightarrow0$}
Before introducing the interaction with the environment, I want to discuss one of the standards conditions that is generally applied to systems in the classical limit. i.e. the condition of negligible quantum action: 
 $$\hbar\rightarrow0$$
 A general critique to this condition is that $\hbar$ is a constant and constants do not go to zero; however, the formula $\hbar\rightarrow0$ is taken as physical shorthand to express the following condition: $$\hbar \ll A_{cl}$$
That is: the quantum action is negligible with respect to the classical actions involved, condition that is usually fulfilled in the macroscopic regime. The inequality $\hbar \ll A_{cl}$ is in fact equivalent to $\frac{\hbar}{A_{cl}} \ll 1$, which becomes, in the macroscopic limit:  $$ \lim_{A_{cl} \to \infty} \frac{\hbar}{A_{cl}}\rightarrow 0$$. 

\noindent The condition $\hbar \rightarrow 0$ is linked to the classical limit since, when this condition is fulfilled, some important features of classical mechanics seem to emerge from the fundamental quantum formalism. For example: 

\begin{itemize}

\item The commutators of quantum mechanics become equivalent to the Poisson brackets of classical mechanics. For example, the canonical commutator  $ \left[ \hat{x},\hat{p} \right]=i\hbar(1) $ in the limit $\hbar \rightarrow 0$ roughly approximates the corresponding Poisson bracket $(x,p)=1 $

\item The discrete energy levels of quantum mechanics approach the continuous energy levels of classical mechanics (high energy/high frequency regime).

\item The quantum Hamilton-Jacobi equation \eqref{qHamilton} seems to approach the classical Hamilton-Jacobi equation, since the quantum potential \eqref{qpotential} explicitly depends on $\hbar$.    

\end{itemize}

\noindent Nevertheless, the problem with this strategy is that it works only for special states.\footnote{Some of these examples are discussed e.g. in Holland (1993, ch. 6).} Many quantum states are completely insensitive to this condition and thus will be unaffected by this strategy. A state of this kind is, for example, the stationary wave function, solution of an infinite potential well. This counter-example was originally proposed by Einstein (1953) to show that quantum mechanics\footnote{And Bohm's theory, briefly discussed by Einstein in the original article} did not lead to the correct classical limit in the expected regime, i.e. the high energy/high frequency regime.\footnote{According to Einstein, those conditions characterized completely the classical limit. The importance of the entanglement with the environment and the resulting decoherence process was not known at that time: the first important works on decoherence were published only some decades later. } 

\noindent Einstein considers the case of a box delimited by two impenetrable walls and a bullet moving to and fro the walls. The idea is that, starting from the fundamental quantum description, the bullet should start moving classically when the classicality conditions are applied (so, when it practically becomes a ``classical'' bullet).  
The quantum description inside the box corresponds to the solution of an infinite potential well, i.e. a stationary wave function:

\begin{equation}
\psi_n(x,t)=\sqrt\frac{2}{L}sin(k_n x)e^{-\frac{i}{\hbar}E_n t} 
\end{equation}

\noindent where $L$ is the length of the box, $\sqrt{2/L}$ a normalization constant and $k_n=n\pi/L$ the wave number of the stationary wave (varying with the different energy levels $E_n$). Einstein notes that this state does not have a classical analogue (the stationary wave--quantum bullet does not become a classical bullet) in the expected classical limit, i.e. the high energy/high frequency regime, conceptually equivalent to $\hbar \rightarrow 0$. On the contrary: in this regime, the stationary wave has an increasing number of nodal points, i.e. points in which the probability to find the particle is zero, and this is the exact opposite of a classical scenario: a particle moving with uniform velocity between the walls, corresponding to a uniform position probability distribution. In this case, applying the classicality conditions, the quantum system  seems to diverge from its classical analogue: higher the energy level considered, higher the wave number and so the number of nodal points inside the box, the more the difference with the description of a classical bullet we wanted to reach applying those conditions.\footnote{This is one of the few paradoxes in quantum mechanics that involves only the classical limit without reference to the collapse of the wave function or the measurement problem. In a way, we could just rename it as the \emph{Einstein's particle-in-a-box} paradox.}

\noindent To see the same problem from a different angle: a stationary wave function is equivalent to a superposition of plane waves with opposite momenta ($p=\hbar k$):

\begin{equation}
\psi(x)=Bsin(kx)=Be^{ikx}+Be^{-ikx}
\end{equation}

\noindent with $B=\sqrt{2/L}$. The stationary wave represents a continuous quantum interference between the two plane waves and this interference is not resolved by the condition $\hbar \rightarrow 0$. Therefore, this state will remain typically quantum even in case the system itself is macroscopic (insofar it is well shielded from the environment). Incidentally, we note that the problem is not alleviated in Bohm's theory,. In this framework, the Bohmian particle remains at rest in one of the non-nodal points inside the box (since $v=\frac{\nabla{S}}{m}=0$) even if the system has a finite (non-zero) total energy and the potential energy inside the box is zero, a highly non-classical situation.\footnote{On this point , see e.g. Holland (1993, ch. 6), Myrvold (2003, sect. 3.1).}

\noindent From the analysis of the Einstein's box problem, it follows that:
\begin{itemize}
\item{The criterion $\hbar\rightarrow0$ alone is not sufficient (or not general enough) for recovering the classical limit of quantum mechanics. There are states, as the ``particle in a box'' state, that are unaffected by this condition;}
\item{The classical limit cannot be generally recovered for isolated systems: the interaction with the environment is a necessary condition to derive (approximately) Newtonian mechanics from the underlying quantum description.}
\end{itemize}

\noindent This does not mean that the condition $\hbar \rightarrow 0$ is not important or not relevant for the classic limit, but that we must apply it to the correct states. Intuitively, the condition of negligible quantum action is equivalent to say that the system is macroscopic (big mass, high energy). But macroscopic systems will likely be open systems, i.e. systems that interact with the surrounding environment. This is indeed a missed element in the Einstein's example:\footnote{Decoherence theory will be formulated only some decades later with respect to Einstein's example.} it is highly improbable that a macroscopic bullet moving between two walls will remain quantum-mechanically isolated for long time, and until it does we do not expect a classical behavior. When the system interacts with the environment, this interaction changes the wave function of the initially isolated system. And since environmental decoherence acts very fast, almost instantaneously for macroscopic systems in ordinary conditions, the wave function of the system will change accordingly very fast, almost instantaneously.\footnote{This claim cannot be taken too literally in the standard interpretation, as we cannot assign a wave function to an open system--subsystem of an entangled state--but only a reduced density matrix. However, it can be taken quite literally in Bohm's theory, where an open system is likely to be described by an effective wave function, especially in the macroscopic regime where there is strong interaction with the environment and the system itself has a large numebr of degrees of freedom.} 

\noindent Therefore, the condition $\hbar \rightarrow 0$ must be applied to the states selected by decoherence: the pointer states in the standard interpretation and the (environmentally-selected) effective wave functions in Bohm's theory. There is indeed an analogy between the two: both are by definition states that remain stable during the interaction with the environment. Using a jargon familiar to the decoherence literature, both are states \emph{selected by decoherence}. After all, we want classical Newtonian mechanics to emerge not for any state, but for those states that are reasonable to expect in the macroscopic regime. And those states are the states selected by decoherence. We can therefore summarize the proposed strategy in three steps:

\begin{enumerate}
\item{The first step is to describe how the wave function of an initially isolated system changes after the interaction with the environment. In Bohm's theory, this interaction produces well-localized effective wave functions (sect. 4, especially subsection 4.3). To be precise, and to stress the difference between those kinds of states and the general effective wave functions in Bohm's theory, we may call the specific states selected by the interaction with the environment \emph{environmentally selected effective wave functions} (ES-EWFs). }
\item{In step 2, I analyze the connection between the ES-EWFs in Bohm's theory and the pointer states of standard decoherence (sections 5.1 and 5.2). This connection will lead us to characterize the ES-EWFs in terms of Gaussian states (sect. 5.3).}
\item{In step 3, I apply the standard classicality conditions ($\hbar \rightarrow 0$ $m \rightarrow \infty$) to the states selected by decoherence, i.e. ES-EWFs mathematically represented by Gaussian states (sect. 6). The analysis shows that, under these conditions, the quantum potential of the system is negligible $\left( Q \rightarrow 0\right)$ and Bohm's theory reduces approximately to Newtonian mechanics.}
\end{enumerate}

\noindent The proposed strategy can thus be summarized by the following table:

\begin{center}
\begin{tabular} {|l|l|l|}
\hline
\textbf{Interaction with $\xi$}  & \textbf{ES-EWFs $\leftrightarrow$ pointer states} & \textbf{macroscopic systems} \\
\hline 
 ES-EWFs (well-localized) & Gaussian states & Newtonian dynamics \\
\hline
\end{tabular}
\end{center}

\section{The emergence of effective wave functions}
The notion of \emph{effective wave function} plays a central role in Bomh's theory. The effective wave function in Bohm's theory is equivalent to the usual wave function in standard quantum mechanics. That is: every time we can assign a wave function to a given system, in Bohm's theory this is an effective wave function. A wave function can be generally assigned to a system as long as this is effectively decoupled from the environment. But how can a given wave function be initially assigned or produced? In standard quantum mechanics, this is generally achieved by a filtering operation, i.e. an interaction between a system and a suitable measurement apparatus that selects the required initial state.\footnote{That is: the selected initial state of a system in standard quantum mechanics is already the result of a measurement process.} In the standard context, the factorization between the system and the environment (including the measurement apparatus) is guaranteed by the collapse of the wave function. The filtering procedure, or any other measurement interaction, makes the wave function of the system to collapse into one of the eigenstates of the measured observable. It is the collapse of the wave function that produces a factorization between the system and the surrounding environment (intuitively: in a measurement interaction, system and apparatus are entangled before the collapse takes place, whereas they are factorized right after the collapse, though the final apparatus state is certainly correlated with the final system state).
In Bohm's theory, instead,  the wave function never collapses and the factorization between system and surrounding environment is realized by a different mechanism known as \emph{effective factorization} (sometimes also called, for this reason, \emph{effective collapse}). The effective factorization thus plays in Bohm's theory the same role of the wave function collapse in quantum mechanics: it permits to assign wave functions to systems.\footnote{See Rovelli (2022) for a recent analysis of the preparation of initial states in Bohm's theory based on the effective factorization process.} However, the effective factorization process can assign a wave function to a system even in cases that have no analogue in standard quantum mechanics, such as, for example, in the case of a system that strongly interacts with the environment. As we will see in the next sections, indeed, effective wave functions are naturally produced through the interaction with the environment, i.e. in the decoherence regime.  


\subsection{Ideal factorization}
First, we consider the case of an ideal or exact factorization, corresponding to the standard product state, In this case, a general N-particle system can be decomposed or factorized into N separate one-particle systems. Consider, for simplicity, a 2-particle system: if we can write the wave function of the total system as a product of two wave functions associated to each particle:

$$\psi_{AB}(x, y)=\psi_A(x)\psi_B(y) $$

\noindent then we can say that the system is \emph{factorizable}. In the factorized state, we can assign a wave function to the relevant subsystems (in this case, each of the two particles). As a consequence, each factorized subsystem is physically independent: each subsystem follows its own Schr\"odinger equation and each particle follows it own guiding equation, as the velocity of particle $X$ or particle $Y$ will depend, respectively, only on $\psi_A$ or $\psi_B$. For example, the guiding equation of particle 1 will be:

\begin{equation}
\frac{dX}{dt}=\frac{\hbar}{m}Im\frac{\nabla_{x}(\psi_A(x)\psi_B(y))}{\psi_A(x)\psi_B(y)}=\frac{\hbar}{m}Im\frac{\nabla_{1}\psi_A(x)}{\psi_A(x)} 
\end{equation}

\noindent where $X$ is the actual position of the Bohmian particle and the right-hand side of the equation is evaluated at $x=X$. In the same manner, the quantum potential acting on particle $X$ will be independent from that of particle $Y$:
\begin{equation}
 Q_A=-\frac{\hbar^2}{2m}\frac{\nabla_x^2 \left( R_A(x)R_B(y) \right)}{R_A(x)R_B(y)} = -\frac{\hbar^2}{2m}\frac{\nabla_x^2  R_A(x)}{R_A(x)}
 \end{equation}

\noindent That is: factorizability implies physical independence. Every time we can decompose a many particle system as a product of N factorized states, these resulting states will be physically independent from one another. In Bohm's theory, this means that the motion (velocity and acceleration) of the particles composing one factorized state (e.g. system A) will not be affected by the position of the particles composing a different factorized state (e.g. system B). In this way, factorizability turns off Bohmian non-locality. 

\noindent However, the condition of ideal factorizability is very fragile and not realistic in the classical limit: any time that initially factorized systems interact with each other through classical potentials, the Schr\"odinger evolution will generally transform the initially factorized state into an entangled state:

\begin{equation}
\psi_A(x)\psi_B(y)\stackrel{H(x,y)}{\rightarrow}\psi_{AB}(x,y)
\end{equation}

\noindent where the entangled state $\psi_{AB}(x,y)$ will be generally a superposition of factorized states:

\begin{equation}
\psi_{AB}(x,y)=\sum_i c_i \psi_i^A(x)\psi_i^B(y) 
\end{equation}
 
\noindent Since the classical macroscopic regime is characterized by continuous interactions between systems, the condition of ideal factorizability will be totally unrealistic in that context. I will take indeed the opposite view: macroscopic Bohmian systems are (effectively) factorized because of the continuous interaction with the surrounding environment. 

\subsection{Effective factorization}

We now consider the case in which the 2-particle system $\psi_(x,y)$ is an entangled state, mathematically represented by the following superposition:

\begin{equation}
\Psi(x,y)=\psi_A(x)\psi_B(y)+\psi_C(x)\psi_D(y) \label{superposition}
\end{equation}

\noindent The system is described by a linear combination of product states, whereas the total state itself is not factorized. In this case, the behavior of particle $X$ and that of particle $Y$ are correlated with each other: if in a measurement we find particle $X$ in the state $A(C)$, then, upon measurement, particle $Y$ will be found with certainty in the state $B(D)$. Still, in some circumstances the superposition \eqref{superposition} becomes equivalent to a factorized state: this happens when the two components of the superpositions have negligible overlap on configuration space:

\begin{eqnarray} 
\braket{\psi_B|y}\braket{y|\psi_D} \approx 0 \quad \forall y \in \mathcal{Q}_Y  \\  \label{overlap1}
\braket{\psi_A|x}\braket{x|\psi_C} \approx 0 \quad \forall x \in \mathcal{Q}_X       \label{overlap2}
\end{eqnarray}

\noindent where $\mathcal{Q}_X$ and $\mathcal{Q}_Y$ are, respectively, the configuration space of particle $X$ and particle $Y$, or, equivalently, when they have approximately disjoint supports in configuration space:

\begin{equation} \label{support}
\textrm{supp} \left(\psi_A(x)\psi_B(y) \right) \cap \textrm{supp} \left( \psi_C(x)\psi_D(y) \right) \approx \emptyset 
\end{equation}

\noindent In this simple case, the configuration space is one-dimensional, so the two particles have negligible overlap when they are separated in three-dimensional space. For an N-particle system, however, the configuration space has $3N$-dimensions and this plays an important role for the condition to be realized. It is sufficient indeed that there is no overlap for only one degree of freedom in order to have no overlap between the two components: higher the number of degrees of freedom, higher the probability that the two components will not overlap (or will negligibly overlap) on configuration space. This point was originally remarked by Bohm \& Hiley (1987, p. 333):

\begin{quote}
This is because of the multi-dimensional nature of the many-body wave function, which implies that the packet $\psi_m(x)\phi_m(y)$ and any other packet, say $\psi_v(x)\phi_v(y)$ will fail to overlap as long as \emph{one} of its factors fails to overlap, even though the other factors would still have some overlap.
\end{quote}

\noindent The condition of no-overlap is a typicality statement: it is a probabilistic statement, but one that is very likely to be realized when systems are macroscopic and composed of approximately $10^{23}$ degrees of freedom (this consideration will come back again later, considering the effective factorization induced by the interaction with the environment). 

\noindent When the two components of the superposition: $\psi_A(x)\psi_B(y)$ and $\psi_C(x)\psi_D(y)$ have negligible overlap on configuration space, they form different branches or different channels, each one separated by a region where the wave function is practically zero and so the probability of finding the particle on that region is also practically zero. As a consequence, once the distinct separated channels are formed, the two particles ($X$ and $Y$) will enter only one channel. This can be intuitively thought as follows: the velocity of the Bohmian particles is proportional to the density of the wave function $|\psi|^2$: so the particles cannot enter into regions where the wave function is practically zero (the region of no-overlap between the different components) while, following the density of the wave, they will enter one of the channels for which the density $|\psi|^2$ is not negligible (higher the density associated to a particular branch, higher the probability that the particle will enter on that branch). Therefore, when the condition \eqref{support} is realized, the initial entangled state \eqref{superposition} becomes equivalent to a proper mixed state, i.e. the superposition represents a state of epistemic ignorance about which one of the two components have been actually implemented or realized by the Bohmian dynamics. The component of the superposition in which the Bohmian particles have been entered into is called \emph{effective wave function}, whereas the other component is called \emph{empty wave function}. If we suppose, for example, that particles $X$ and $Y$ entered into the first component,\footnote{Note that, in this process, the dynamics of the two particles is correlated: if particle X enters the component A(C), particle Y will necessarily enter the component B(D).} then the initially entangled state \eqref{superposition} can be written as follows:

\begin{equation}
\Psi(x,y)=\psi^{eff}_A(x)\psi^{eff}_B(y)+\psi_C^{emp}(x)\psi_D^{emp}(y)
\end{equation}

\noindent where $\psi^{eff}_A(x$ and $\psi^{eff}_B(y)$ are, respectively, the effective wave functions of particle $X$ and particle $Y$ and $\psi_C^{emp}(x)$, $\psi_D^{emp}(y)$ are the empty wave functions. From this moment on, the state of the Bohmian system is completely represented by the product state of the effective wave functions, whereas the empty wave functions can be discarded.\footnote{The empty components are discarded for practical purposes but, if they happen to overlap with the effective wave function in the future evolution of the system, then they become relevant again and produce interference. For this reason, Bohm \& Hiley (1987) preferred to call them ``inactive'' rather than empty components. } 
In short, when the condition \eqref{support} is realized, the entangled state \eqref{superposition} is practically equivalent to the following factorized state:

\begin{equation}
\Psi(x,y) \approx \psi_A(x) \psi_B(y)
\end{equation}

\noindent  This is the process of effective factorization: one of the components of the initial superposition is selected as the effective state and this will be the effective wave function. Sometimes this process is also called \emph{effective collapse}, as the Bohmian system practically behaves as it had collapsed into one of the two components of the initial superposition.\footnote{The process of effective factorization has been originally described by Bohm \& Hiley (1987) and the notion of effective wave function appears in sect. 7 (p. 344)). A detailed analysis of the effective factorization is also given by D\"urr et al. (1992), D\"urr \& Teufel (2009, ch. 9)), Holland (1993, ch. 8).}

\noindent Effectively factorized states are generally produced by measurement interactions, i.e. interactions between a system (whose initial state is a superposition of eigenstates of the measured observable) and a measurement apparatus, as the different pointer states of the apparatus correspond to macroscopically distinct states, typically represented by non-overlapping states on configuration space. Consider, for example, the interaction between a system, represented by the following superposition:

$$\psi(x)=\sum_ic_i\psi_i(x) $$

\noindent where $\psi_i$ are the eigenfunctions of the measured observable, and the measurement apparatus $\phi(y)$, represented initially by the ready state $\phi_0(y)$. The global initial state is a factorized state of the system, $\psi(x)$, and the apparatus, $\phi_0(y)$, i.e $\Psi(x,y)=\psi(x)\phi_0(y)$. During the measurement interaction, the Schr\"odinger evolution transforms the initial factorized state into an entangled state, i.e. a superposition of correlated system and apparatus relative states:

$$\psi(x)\phi_0(y) \rightarrow\sum_{i}c_i\psi_i(x)\phi_i(y) = \Psi(x,y) $$

\noindent If the different pointer states of the apparatus $\phi_i(y)$ have (approximately) disjoint supports on the apparatus' configuration space, then the final state $\sum_{i}c_i\psi_i(x)\phi_i(y)$ reduces to an effectively factorized state,\footnote{On this point, see e.g. Rovelli (2022), in particular eqs. (5) and (6).} i.e. a state practically equivalent to an effective wave function and many empty wave functions.\footnote{Effectively factorized components with no Bohmian particles.} Suppose, for example, that the measurement result is the eigenvalue $s$ associated to the eigenstate $\phi_s(y)$: the (total) effective wave function in this case will be the factorized state $\psi_s(x)\phi_s(y)$, where $\psi_s(x)$ and $\phi_s(y)$ are, respectively, the effective wave functions of the system and the apparatus:

\begin{equation}
\Psi(x,y)=\sum_i c_i \psi_i(x)\phi_i(y) \approx \psi_s(x)\phi_s(y)
\end{equation}

\noindent Since the effective or effectively collapsed component is a factorized state: $\psi_s(x)\phi_s(y)$, the system and the apparatus will be dynamically independent after the measurement interaction. The formation of effective wave functions for the subsystems of an initially entangled state correspond to subsystems that are dynamically independent from each other. As a consequence, at regimes where the formation of effective wave functions is very fast and widespread, we expect the typical Bohmian non-locality to be completely turned off.\footnote{See e.g. Bohm \& Hiley (1993, sect. 7.6). }. This is indeed what we expect in the macroscopic regime, where the environment plays the same role of the measurement apparatus in the process of effective factorization.   



\subsection {Effective factorization through environmental interaction}

The process of effective factorization in Bohm's theory is thus generally associated with measurement interactions:\footnote{See e.g. by D\"urr et al. (1992, p. 24-25).} different pointer states of the apparatus will tend to have (approximately) disjoint supports on configuration space, given the large number of degrees of freedom of a macroscopic apparatus and that the different pointer states are macroscopically distinct.\footnote{See e.g. Romano (2016, p. 13).} Yet, there is another situation in which the process of effective factorization is generally realized, that is, when the system interacts with the surrounding environment. The latter, as the measurement apparatus, is a system composed of a large number of particles (proportional to the Avogadro's number) that interacts continuously with the system.\footnote{There is a difference though between the two: the apparatus interacts with the system at a given time and all the apparatus' degrees of freedom are involved on that interaction at that single time, whereas the environment is generally described as a tensor product of many particles which interact one at a time with the system. However, these two descriptions eventually converge to the same result, with the difference that the decoherence process induced by the environment is continuous and progressive whereas that one induced by the measurement apparatus is instantaneous and discrete (this corresponds to the collapse of the wave function in the standard interpretation). } In order to stress the analogy between the two, in the decoherence literature it is often said that the system is ``monitored'' or even ``measured'' by the environment. This is particularly true in Bohm's theory, where the environment acts literally as a macroscopic measurement device. In Bohm's theory, indeed, there is no difference between decoherence and measurement: both are interactions with a macroscopic external system that, amplifying the number of degrees of freedom of the total system, leads dynamically to a separation of the entangled system-environment (or system-apparatus) wave function into non-overlapping components. We can schematically represent the interaction between an entangled 2-particle system and an external system (the ``environment'') as follows:

$$\Psi(x,y)(\Xi_1(z_1)\dots\Xi_N(z_N))=\sum_{i}\psi_i(x)\phi_i(y)(\Xi_1(z_1)\dots\Xi_N(z_N))\stackrel{H_{int}}\rightarrow\sum_{i}\psi_i(x)\phi_i(y)\xi_i(z_1,\dots,z_N)$$   

\noindent The Bohmian system will be effectively factorized--or decohered--when the different relative environmental states have disjoint supports on configuration space:

\begin{equation} 
\textrm{supp} \left( \xi_m(z_1,\dots,z_N)  \right) \cap \textrm{supp} \left( \xi_n(z_1,\dots,z_N)  \right) \approx \emptyset 
\end{equation}

\noindent The effective factorization of the wave function, leading to the formation of the effective wave functions for the subsystems, is the mechanism that explains the loss of quantum non-locality in Bohm's theory. This point was originally stresses by Bohm \& Hiley (1987, p. 344):

\begin{quote}
Quantum non-local connection is fragile and easily broken by almost any disturbance or perturbation. [\dots] we may expect that non-local connection will not normally be encountered under ordinary conditions, in which every system is bathed in electromagnetic radiation and is subject to external perturbations of all kinds as well as random thermal energies. 
\end{quote}

\noindent In short, the effective factorization process--leading to the formation of the effective wave functions--explains why Bohmian non-locality is not manifest at the macroscopic regime, where the interaction with the environment becomes very effective. This is the physical basis of decoherence in Bohm's theory: the (dynamical, progressive) formation of effective wave functions leads to the impossibility to detect quantum interference between different subsystem's components. Decoherence is an empirical effect, the effective factorization is its physical explanation in terms of the wave function global behavior. We therefore reach the conclusion that the effective wave functions are primarily produced by the interaction with the environment. But which kind of wave function should we expect to be formed by the interaction with the environment? This question will be addressed in the next section.

\section{Connecting Bohm's theory with decoherence}

\subsection{Pointer states}
One of the main results of quantum decoherence is the selection of a special kind of states for the subsystem in interaction with the environment which remain stable in the decoherence process. Such states are called \emph{pointer states}: they are practically immune to decoherence, as they do not get entangled (or they get least entangled) with the environment. The standard definition of pointer states--or \emph{pointer basis}, when the pointer states form a complete basis in the subsystem's Hilbert space--is due to Zurek (1981, 1982). He called such states ``pointer states'' as, historically, the interaction with the environment was used to explain the stability of the different macroscopic states of a measurement device in a measurement process. That is: the environment interacts with the macroscopic apparatus and selects definite macroscopic (pointer) states, while superpositions of such states quickly decohere. The pointer states are the only states that one can observe after decoherence: 

\begin{quotation}
[...] we may say that the interaction with the environment superselects the observable states of the system: Some states are robust in spite of the environmental interaction, while other states are rapidly decohered and become therefore unobservable in practice. However, in contrast with the postulated exact superselection rules, environment-induced superselection represents effective superselection rules that dynamically emerge from
the (structure of the) system--environment interaction. [Schlosshauer (2007, p. 73)]
\end{quotation} 

\noindent The pointer states of the system are those state that remain stable during the interaction with the environment, i.e. those state that get least entangled with the environment. Consider, for example, a system represented by a superposition of two states:

\begin{equation}
\ket{\psi_S}=c_1\ket{S_1}+c_2\ket{S_2} \label{system}
\end{equation}

\noindent interacting with the external environment $\ket{\xi}$ via the interaction Hamiltonian $\hat{H}_{int}$. The system states that remain stable during the interaction will be the eigenstates of the interaction Hamiltonian:

\begin{equation}
\ket{\psi_S}\ket{\xi}=\left( c_1\ket{S_1}+c_2\ket{S_2} \right) \ket{\xi} \stackrel{\hat{H}_{int}}\longrightarrow c_1\ket{S_1}\ket{\xi_1}+c_2\ket{S_2}\ket{\xi}_2  \label{interaction}
\end{equation}

\noindent The dynamics that selects pointer states can be schematically described as follows:

\begin{equation}
e^{\frac{i}{\hbar}H_{int}t}\ket{S_i}\ket{\xi_0}=\lambda_i\ket{S_i}e^{\frac{i}{\hbar}H_{int}t}\ket{\xi_0} \equiv \lambda_i \ket{S_i}\ket{\xi_i(t)} \label{pointer}
\end{equation}

\noindent where $i=1,2$ in the example \eqref{interaction},  $\ket{S_i}$ are eigenstates of the interaction Hamiltonian and $\lambda_i$ are the eigenvalues associated to the eigenstates $\ket{S_i}$. In the macroscopic regime, the interaction between the system and the environment is generally very strong. So strong that the interaction Hamiltonian ($\hat{H}_{int}$) generally dominates over the other terms, i.e. the self-Hamiltonian of the system ($\hat{H}_{S}$) and that one of the environment $(\hat{H}_{\xi}$):

\begin{equation}
\hat{H}_{int} \gg \hat{H}_{S}  + \hat{H}_{\xi}
\end{equation}

\noindent and so we can represent the dynamics as driven by the interaction Hamiltonian. Also, as the interaction Hamiltonian is generally represented by a function of the position coordinates (typically: classical potentials or scattering), the system pointer states will be approximate position eigenfunctions, i.e. well-localized states in the position basis.\footnote{See e.g. Schlosshauer (2019, pp. 8-10).}\\
However, in more realistic cases the pointer states cannot be found using \eqref{pointer} as different components contribute to the total dynamics and the pointer states are not exact eigenstate of the interaction Hamiltonian. This is the case, for example, of the quantum Brownian motion, which is a decoherence model typically used in the macroscopic regime. More indirect methods have been developed to determine the approximate pointer states of the system in those situations, such as Zurek's predictability sieve.\footnote{Zurek (1993).}. In this case, the system pointer states will not be states in a proper sense, as they cannot be represented by a state vector or a wave function, but only by a reduced density matrix. As we will see, the process of effective factorization will be of great help: while in quantum mechanics we are forced to recognize that the subsystems pointer states are not exactly ``states'', the effective factorization process will be able to identify those states as \emph{environmentally-selected effective wave functions} (ES-EWFs). This will provide a clear interpretation of Zurek's result on the selection of pointer states in the quantum Brownian motion in terms of ES-EWFs, and it will be analyzed in section 5.3.2.

\subsection{ES-EWFs are pointer states}
The analogy between the ES-EWFs in Bohm's theory and the pointer states in standard decoherence theory emerges when we compare the process that originates the former and the latter in the respective frameworks. Let us then analyze and compare the two descriptions, starting with decoherence theory. The starting point of decoherence is the interaction--mathematically represented by a suitable interaction Hamiltonian--between a system and an external system, called the environment. Furthermore, the system initial state has to be a superposition of  states, as in \eqref{system}, in order for the system to get entangled with the environment. The final state is thus a system-environment entangled state:  

\begin{equation}
\ket{\psi_s}\ket{\xi}=\left(c_1\ket{S_1}+c_2\ket{S_2} \right)\ket{\xi}\stackrel{H_{int}}\longrightarrow \alpha \ket{S_{\alpha},\xi_{\alpha}} + \beta \ket{S_{\beta},\xi_{\beta}}  
\end{equation}

\noindent This is valid in the general case, i.e. for any state $\ket{S_i}$ of the system and for any interaction $H_{int}$. The pointer states however are special states: they remain stable under the action of $H_{int}$ and get perfectly correlated with the relative states of the environment after the interaction (as described before, this happens when the system states are eigenstates of the interaction Hamiltonian operator):

\begin{equation}
\ket{\psi_S}\ket{\xi}=\left( c_1\ket{S_1}+c_2\ket{S_2} \right) \ket{\xi} \stackrel{\hat{H}_{int}}\longrightarrow c_1\ket{S_1}\ket{\xi_1}+c_2\ket{S_2}\ket{\xi}_2 = \ket{\Psi} \label{superposition2}
\end{equation}

\noindent The final state $\ket{\Psi}$ is an entangled state between the system and the environment. However, in decoherence we are generally interested in the behavior of the subsystem $S$ in interaction with the environment $\ket{\xi}$. Consider, for example, a table ($\ket{S}$) scattered off by air molecules ($\ket{\xi})$: what we want to know is how the behavior of the environment (air molecules) affects our system of interest (the table). But we are  faced with a problem: the system (the table, in our example) does not have its own wave function  after the interaction with environment, as this is a subsystem of a larger entangled state. This problem is solved by the reduced density matrix (RDM), which can be assigned to the subsystem tracing out the degrees of freedom of the environment:

\begin{equation}
\rho_S=Tr_{\xi} (\rho_{\Psi}) \label{reduced_S}
\end{equation}

\noindent The RDM \eqref{reduced_S} describes the probability distribution of all possible measurement outcomes of a measurement performed on the subsystem $S$. Considering the final entangled state \eqref{superposition2}, the subsystem RDM takes the following form:

\begin{equation}
\rho_S = |c_1|^2 \ket{S_1}\bra{S_1} \braket{\xi_1|\xi_1}+ c_1c_2^{*}\ket{S_1}\bra{S_2}\braket{\xi_1|\xi_2}+ c_1^{*}c_2 \ket{S_2}\bra{S_1}\braket{\xi_2|\xi_1}+ |c_2|^2 \ket{S_2}\bra{S_2} \braket{\xi_2|\xi_2} \label{density}
\end{equation}

\noindent Assuming the condition of orthogonality of the environmental states:\footnote{This is an assumption in decoherence theory, but it is possible to show that this condition is fulfilled in the long run (i.e. after many interactions between the system and the external particles composing the environment) even if the different relative environmental states are not orthogonal.}

\begin{equation}
\braket{\xi_1|\xi_2}=0 \quad,\quad \braket{\xi_2|\xi_1}=0 \label{orthogonal}
\end{equation}  

and the standard normalization of states:
\begin{equation}
\braket{\xi_1|\xi_1}=1 \quad,\quad \braket{\xi_2|\xi_2}=1
\end{equation}

eq. \eqref{density} reduces to

\begin{equation}
\rho_S = |c_1|^2 \ket{S_1}\bra{S_1} + |c_2|^2 \ket{S_2}\bra{S_2}  \label{reduced_density}
\end{equation}

\noindent That is: if the different relative environmental states are orthogonal with each other, then the subsystem's reduced density matrix is diagonalized, and the diagonal terms of the RDM are the pointer states selected by the interaction with the environment. From eq.\eqref{reduced_density} we see that the states $\ket{S_1}$ and $\ket{S_2}$ are the pointer states. \\
We are now in the position to compare the selection of the pointer states in decoherence theory with the formation of the effective wave functions through the interaction with the environment (ES-EWFs) in Bohm's theory. The starting point, as in the case of decoherence, is the interaction between a system and the external environment, with the system initial state being described by a superposition of states. Differently from standard decoherence, however, Bohm's theory privileges a position-basis representation.\footnote{Position is a privileged quantity in Bohm's theory. For example: the velocity of the Bohmian particles is given by the gradient of the phase of the wave-function in the position representation.} For this reason, it is more convenient to translate the process in the wave function representation:

\begin{equation}
\psi(x) \xi(y)=\left( c_1\psi_1 (x)+ c_2 \psi_2(x) \right) \xi(y) \stackrel{H_{int}(x,y)}\longrightarrow \alpha \phi_1(x,y) + \beta \phi_2(x,y)  \label{superposition3}
\end{equation}

\noindent for general states $\psi_i(x)$ and a general system-environment interaction $H_{int}(x,y)$. However, if the system states $\psi_i(x)$ are eigenfunctions of the interaction Hamiltonian operator:

$$\psi_1(x )\xi(y) \stackrel{H_{int}(x,y)} \longrightarrow \lambda_1\psi_1(x)\xi_1(y) $$
$$\psi_2(x )\xi(y) \stackrel{H_{int}(x,y)} \longrightarrow \lambda_2\psi_2(x)\xi_2(y) $$

the final state of eq. \eqref{superposition3} can be represented by the following superposition:

\begin{equation}
\psi(x) \xi(y)=\left( c_1\psi_1 (x)+ c_2 \psi_2(x) \right) \xi(y) \stackrel{H_{int}(x,y)}\longrightarrow c_1 \psi_1(x)\xi_1(y) +  c_2 \psi_2(x) \xi_2(y) =\Psi(x,y)  \label{decomposition}
\end{equation}

\noindent Eq. \eqref{decomposition} is equivalent to eq. \eqref{superposition2}: it represents the interaction between the system and the environment and the final.entangled system-environment state, just translated in the wave function representation. As we saw before, in order to select the subsystem pointer states from the final entangled state in decoherence theory, we need to introduce the condition of orthogonality of the environmental states. This condition indeed leads to the diagonalization of the RDM and the diagonal terms of the matrix are the pointer states. But this is exactly the condition that we need in Bohm's theory to select the effective wave functions, just translated in the position-basis representation\footnote{The condition of superorthogonality between environmental states or, equivalently, the requirement of disjoint supports, is the implementation of the orthogonality condition in the position basis.}. That is: as the subsystem states $\ket{S_1}$, $\ket{S_2}$ are pointer states if the relative environmental states correlated to those states are orthogonal: 

\begin{equation} \label{ortogonale}
\ket{S_1} \textrm{,}  \ket{S_2} \quad \textrm{are pointer states if} \quad \braket{\xi_1|\xi_2}=0
\end{equation}

\noindent in Bohm's theory the subsystem states $\psi_1(x)$ and $\psi_2(x)$ are effective wave functions if the relative environmental states correlated to those states have disjoint supports on configuration space:

\begin{equation} \label{supporto}
\psi_1(x) \textrm{,} \psi_2(x) \quad \textrm{are effective wave functions if} \quad \braket{\xi_{1(2)}|y}\braket{y|\xi_{1(2)}} \approx 0 \textrm{,} \quad \forall y \in \mathcal{Q}_{\xi}
\end{equation}

\noindent Comparing \eqref{ortogonale} and \eqref{supporto}, we note that the condition of disjoint supports is essentially the condition of orthogonality of states implemented in the position basis representation.\footnote{The condition for the effective factorization--disjoint supports on configuration space-- is stronger than the standard decoherence condition--orthogonality of states. Nevertheless, we expect the condition of disjoint supports to be approximately or exactly satisfied when the number of degrees of freedom of the environment is very large and the system-environment interaction very strong, as it happens at the macroscopic regime.} \\

\noindent {Eqs. \eqref{decomposition} and \eqref{superposition2} are different representations of the same idea: if the interaction--and correlation--between the system and the environment leads to the formation of non-overlapping components in the Hilbert space (decoherence theory) or on configuration space (Bohm's theory), then the subsystem's relative states will be separated and distinguishable. In decoherence theory, this amounts to say that we cannot detect quantum interference between the different relative states in a ``local'' measurement on the subsystem. In Bohm's theory, we can say something more: the dynamics of the Bohmian particles will select only one of the subsystem's relative states and so we will be able to assign a new wave function to the selected component. This will be eventually the effective wave function.\\
Since in Bohm's theory the notion of effective wave function is very general (basically equivalent to the notion of the wave function in standard quantum mechanics), it is better to call \emph{environmentally-selected effective wave functions} (ES-EWFs) the effective wave functions produced by the interaction with the environment.
The effective factorization process in Bohm's theory is thus analogous to the diagonalization process of the reduced density matrix in decoherence theory. In the latter, we use the density matrix representation as we cannot assign wave functions to subsystems of a larger entangled state (there is no effective collapse in quantum mechanics during the linear Schr\"odinger evolution). In the former, instead, the dynamics of the Bohmian particles permits to re-assign a wave function to the subsystems of a lager entangled state, when the entangled state is so decomposed that the different components do not overlap--or negligibly overlap-- on configuration space. Once this analogy between decoherence in quantum mechanics and the effective factorization in Bohm's theory is clearly spelled out, we recognize the equivalence between the ES-EWFs and the decoherence's pointer states.

\subsection{ES-EWFs are Gaussian states}
But what kind of wave functions are the ES-EWFs? Given the equivalence between the ES-EWFs and the pointer states, we can use some important results about pointer states from the decoherence literature to characterize the ES-EWFs in Bohm's theory.  We will reach the same conclusion through two different arguments: 

\begin{enumerate}
\item{A qualitative example, based on the behavior of the RDM in collisional decoherence;}
\item{A rigorous example, based on Zurek et al. (1993) derivation of the pointer states in the quantum Brownian motion (QBM).}
\end{enumerate} 

\subsubsection{Pointer states from collisional decoherence}
The RDM $\rho_S (x,x',t)$ of a system in interaction with the environment is a function of the system evaluated at two different points, generally indicated as $x$ and $x'$. It thus has a natural representation on a plane, with diagonals $x=x'$ and $x=-x'$, representing two characteristic parameters of the matrix: the \emph{ensemble width} and the \emph{coherence length}, respectively. The ensemble width (corresponding to the quantum probability density $|\psi|^2$), indicates all the possible locations in which the subsystem can be found in a position measurement. The coherence length corresponds to the region over which we can detect quantum interference between different subsystem's relative states. The dynamics of the RDM, mathematically represented by the master equation, leads generally to a spread of the RDM in the direction of the ensemble width and a contraction of the RDM in the direction of the coherence length. This process is the so-called diagonalization of the RDM: higher the contraction in the direction of the coherence length, more difficult is to detect quantum interference over spatially distant subsystem's relative states. A typical master equation for collisional decoherence\footnote{Collisional decoherence is decoherence induced by scattering of environmental particles on the system of interest. It is thus a type of environmental decoherence and one of the most important models of decoherence for the quantum-to-classical transition, together with the Quantum Brownian Motion.} in the long wavelength limit\footnote{This limit applies when the wavelength of the environmental particles is larger than the spatial distance between the subsystem components. } is:\footnote{See Schlosshauer (2007, p. 128-132) for the derivation of the master equation.}

\begin{equation} \label{master}
\rho_S(x,x',t)=\rho(x,x',t_0)e^{\Lambda (x-x')^2 t}
\end{equation}    

\noindent The master equation \eqref{master} shows that decoherence will be very effective for spatially distant points $|x-x'|$ of the subsystem\footnote{Note: the distance $|x-x'|$  has a threshold after which the long-wavelength limit is not valid anymore and has to be replaced by the short-wavelength limit, which applies when the wavelength of the environmental particles is shorter than the spatial distance $|x-x'|$. Decoherence in this latter limit is much more effective than decoherence in the long-wavelength limit.} and, even for small distances, there is an exponential increase of decoherence over time--in particular, the term $D=\Lambda(x-x')^2$ can be thought of as a decoherence rate and, accordingly, the term $\tau_D=D^{-1}$ as a decoherence time. This dynamical process leads to an asymptotic diagonalization of the RDM. In this asymptotic state, the coherence length of the system will be approximately zero, and this is generally interpreted as a signal that the subsystem must be represented by a well-localized state. This is indeed the only possible state which is compatible with this situation, in which quantum interference cannot be detected if not over a very small distance. From the analysis of the master equation in collisional decoherence we thus reach the following conclusion: a subsystem in interaction with the environment will be described by a well-localized state.\footnote{This is a standard claim in decoherence theory. However, see Romano (2022) for a critical assessment of this claim. }

\subsubsection{Pointer states from the quantum Brownian motion}
A rigorous derivation of the specific form of the pointer states is given by Zurek \emph{et al.} (1993) in the framework of the Quantum Brownian Motion (QBM), which is (together with collisional decoherence) one of the most important models of environmental decoherence for the quantum-to-classical transition. In the QBM, the system is represented by a quantum harmonic oscillator and linearly coupled through position with a thermal bath of harmonic oscillators, i.e. a collection of individual quantum harmonic oscillators at constant temperature $T$. In the following, I will briefly resume the strategy used by Zurek and collaborators to derive the specific form of the pointer states.\\As a first step, the authors derive the subsystem reduced density matrix $\rho_S$ and the corresponding mater equation $\dot{\rho}_{S}$, defining the linear quantum entropy $\mathcal{S}$ as a function of $\rho_{S} $:\footnote{The explicit form of the master equation is not relevant for our discussion, but for the interested reader this corresponds to eq. (1) of the original article. A complete analysis of the Quantum Brownian Motion can be found in Schlosshauer (2007, sect. 5.2).}

\begin{equation}
\mathcal{S}(\rho_S) = Tr \left( \rho_{S}-\rho_{S}^{2} \right) 
\end{equation}

\noindent The linear entropy describes the level of purity of a state: for a pure state $(\rho = \rho^{2})$, the linear entropy is zero. It increases when the pure state transforms into a mixture and is maximum for a genuine mixed state. Recalling that pointer states are the states that get least entangled with the environment, i.e. they remain as ``pure'' as possible, they will correspond to those initial states for which the increase of linear entropy is minimum.\footnote{There is an analogy that can be drawn between pointer states in quantum mechanics and material points in classical mechanics. Since the pointer states are those states that get least entangled with the environment, they  follow a quasi reversible dynamics and so they are considered to mimic the reversible dynamics of a material point in classical mechanics. In standard quantum mechanics, a pointer states is thus the closest quantum analogue of a material point in classical mechanics.}  The authors find that the increase of the linear entropy is given by the following expression: 

\begin{equation}
\mathcal{\dot{S}}(\rho_{S}) = 4 D \Delta x^2 
\end{equation}

\noindent where $\Delta x$ is the standard deviation of position and $D=\frac{2 \gamma m k_B T}{\hbar^2}$ is a constant of the model, with $\gamma$ representing the coupling with the environment, $m$ the mass of the system, $k_B$ the Boltzmann's constant and $T$ the temperature of each individual harmonic oscillator composing the environment. In particular, the authors analyze the problem in the weak-coupling limit,\footnote{This permits to describe an approximately reversible dynamics for the pointer states that could be irremediably lost in the strong coupling limit, where the interaction Hamiltonian dominates over the other terms and the subsystem dynamics becomes quickly irreversible.} i.e. the regime in which the self Hamiltonian of the system and the interaction Hamiltonian simultaneously affect the system's dynamics. For mathematical convenience, the authors rewrite the linear entropy as a function of $\Delta x$, $\Delta p$ and $w$, respectively the standard deviations of position and momentum and the angular frequency of the system. Finally, they find that the linear entropy increase is minimum for an initial state characterized by the minimum uncertainty $\Delta x \Delta p = \frac{\hbar}{2}$ and:

\begin{equation}
 \Delta x ^2 = \frac{\hbar}{2mw} 
 \end{equation}
 
\noindent which is the spread in position of the ground state of the quantum harmonic oscillator. From this result, the authors conclude that the pointer state of the QBM  is the ground state of the quantum harmonic oscillator, i.e. a minimum-uncertainty \emph{Gaussian state}.\footnote{The conclusion that the pointer states selected by decoherence are Gaussian states has been recently reinforced by the results of Diosi \& Kiefer (2000) and S\"orgerl \& Hornberger (2015). }

\subsubsection{Pointer states are not states, but ES-EWFs are}
The problem with this conclusion is that, strictly speaking, the ground state of the harmonic oscillator is not there, as there is no state for the subsystem. The form of the pointer state has been inferred by the spread of the position and momentum computed through the reduced density matrix and its time evolution. But there is no state vector or wave function associated to the ground state of the harmonic oscillator. We recall, indeed, that according to standard quantum mechanics the state of a system is completely represented by the state vector, or equivalently by the wave function, while density matrices have been originally introduced in the quantum formalism for computational purposes. Without a real collapse of the entangled system-environment state induced by the interaction with a measurement apparatus, quantum mechanics is just not able to assign a state vector or a wave function to the subsystem of a larger entangled state.\footnote{I have analyzed this point in more detail in Romano (2022, sect. 1.2).} Consequently, the subsystem's pointer states are not properly states in standard quantum mechanics.  
This is a general situation in decoherence theory: the form of the pointer states is generally inferred from specific considerations on the behavior, the dynamics or the structure of the reduced density matrix.\footnote{A different approach is taken by Sorgel \& Hornberger (2015): in their strategy, the pointer states are represented by soliton-like solutions of the system-environment entangled wave function. Even if different from Zurek's approach, also this strategy amounts to individuate some stable dynamical structures within the wave function and reify them as states. So, it does not seem to alleviate the problem discussed here.} This is fine for computational purposes, but it seems more problematic if we want to derive ontological conclusions about the nature of pointer states. Even if the reduced density matrix suggests that the dynamics of the subsystem is such ``to look like'' or ``to behave as'' a Gaussian state, it would be wrong to think that we can \emph{ipso facto} assign a genuine Gaussian state to the subsystem.\\ 
This problem is solved in the context of Bohm's theory: not only we can identify the pointer states with the ES-EWFs produced by the interaction with the environment, but the ES-EWFs are genuine states in Bohm's theory. In Bohm's theory, we can safely re-assign an effective wave function to a subsystem as soon as an effective collapse is dynamically realized, i.e. as soon as the dynamics of the system-environment entangled wave function evolves into negligibly-overlapping components on configuration space. And this is exactly the case when a Bohmian system interacts with the external environment, forming ES-EWFs for the subsystems. We can thus identify the pointer states selected in decoherence theory as ES-EWFs in Bohm's theory. Since the pointer states selected in the macroscopic regime--as shown by Zurek's et al. (1993) for the QBM--look like or behave as Gaussian states, we can simply draw the line and make the obvious conclusion: ES-EWFs in Bohm's theory are Gaussian states.\footnote{For example: in the case of the QBM analyzed by Zurek et al. (1993), the ES-EWF of the system is the ground state of the quantum Harmonic oscillator.}

\section{The emergence of Newtonian dynamics}
In the previous sections I showed that, in Bohm's theory, the interaction with the environment makes the system to collapse effectively (i.e. dynamically) into a well-localized state, in particular--in the macroscopic regime--into a Gaussian state. I called such states \emph{environmentally-selected effective wave functions} (ES-EWFs). The ES-EWFs can be identified with the pointer states of standard decoherence: they are states (effective wave functions) that are dynamically selected by the interaction with the environment. This process of effective factorization explains two features that we expect to occur in the classical limit:

\begin{enumerate}
\item The wave function of the system is well-localized. For a macroscopic system, such as a measurement apparatus, a cat or a table, the wave function is well-localized around the center of mass of the system;
\item Bohmian non-locality drops off progressively yet very quickly: systems described by different ES-EWFs will be quantum mechanically separated, making quantum non-locality disappear in the macroscopic classical regime.  
\end{enumerate} 

\noindent There is however a further important feature that characterizes classical systems: they move according to Newton's theory. That is: the Bohmian dynamics must approximately\footnote{Within a range of approximation that is not empirically detectable at the macroscopic scale.} reduce to the Newtonian dynamics when systems are ``big enough''.  Indeed, we do not expect decohered systems to follow the Newtonian dynamics at every scale, but we do expect the Newtonian dynamics to emerge when the standard macroscopic conditions are realized: 

\begin{itemize}
\item $m \rightarrow \infty$
\item $\hbar \rightarrow 0$
\end{itemize}

\noindent We thus expect that the Bohmian dynamics approximately reduces to the Newtonian dynamics for ES-EWFs (providing the condition of decoherence and interaction with the environment) to which the conditions $m \rightarrow \infty$ and $\hbar \rightarrow 0$ apply. The transition from the Bohmian to the Newtonian dynamics is straightforwardly illustrated within the Hamilton-Jacobi formalism. As we saw in sect. (2), the dynamics of the Bohmian systems is governed by the quantum Hamilton-Jacobi equation: 

\begin{equation}
\frac{\partial{S}}{\partial{t}}+\frac{(\nabla{S})^2}{2m}+V+Q=0 \label{qqHamilton}
\end{equation}

\noindent representing particles with momentum $p=\nabla{S}$ and affected by the (classical + quantum) potential $V+Q$. When the quantum potential is negligible:

 $$Q \approx 0$$, 
 
\noindent eq. \eqref{qqHamilton} reduces to the classical Hamilton-Jacobi equation:

\begin{equation}
\frac{\partial{S}}{\partial{t}}+\frac{(\nabla{S})^2}{2m}+V=0
\label{cHamilton}
\end{equation}

\noindent representing particles with momentum $p=\nabla{S}$ and affected by the classical potential V. Eq. \eqref{cHamilton} corresponds to the fundamental equation of classical Newtonian dynamics, i.e. Newton's second law of motion:

\begin{equation}
m\ddot{x}=-\nabla{V} \label{classicalNewton}
\end{equation}

\noindent Therefore, Bohm's theory reduces to Newton's theory when the quantum potential is zero, or approximately zero, insofar it is responsible for small deviations from the usual Newtonian behavior that are not empirically detectable in the macroscopic regime.  

\noindent {One important note: if we use the quantum Newton's law instead of the quantum Hamilton Jacobi equation we may get the impression that, for the classical limit, not only the quantum potential but also the quantum force (the gradient of the quantum potential) must be zero in order for Bohm's theory to reduce to Newton's theory. This is because in normal conditions the acceleration of the Bohmian particles is affected by the quantum force. However, when the quantum potential is zero or approximately zero, \eqref{qqHamilton} reduces to \eqref{cHamilton} and the quantum force term will not appear in the quantum Newton's law.\footnote{We may think of extreme cases in which, even if the quantum potential has a very small value, its spatial derivative oscillates very rapidly and produces of a non-negligible quantum force. Such cases cannot be excluded a priori, but for the moment we can (quite safely) assume that when the quantum potential is approximately zero the influence of the quantum force will be also negligible.}  

\noindent We must thus analyze the behavior of the quantum potential for the effective wave functions selected by the interaction with the environment. As we saw in sect. (5.3), the ES-EWFs are Gaussian states. A simple Gaussian state can be described as follows:

\begin{equation}
\psi(x)=Ae^{-\frac{x^2}{4\sigma^2}}  \label{gaussian}
\end{equation}

\noindent where $A$ is the amplitude and $\sigma$ the standard deviation of the Gaussian. In order to study the dynamics of the system and compare it with Newton's theory, we compute the quantum potential $Q=- \frac{\hbar^2}{2m}\frac{\nabla^2{R}}{R}$, which encodes the possible deviations from Newtonian dynamics. The quantum potential for the Gaussian state \eqref{gaussian} is given by:\footnote{Ballentine (2015, ch. 14).}
 
 \begin{equation} 
 Q_G=\frac{\hbar^2}{4m\beta^2}(1-\frac{x^2}{\beta^2}) 
 \label{qgauss}
 \end{equation}
 
 \noindent where $\beta^2=\sigma^2[1+(\frac{\hbar t}{2m\sigma^2})^2]$ and $m$ is the mass of the system. It is easy to see that \eqref{qgauss} tends to zero for the standard macroscopic conditions ($m \rightarrow \infty$, $\hbar \rightarrow 0$):\footnote{Note that the quantum potential in \eqref{limit} goes to zero even if we maintain $\hbar$ as a constant and let the limit varying on the value of the mass, which approaches $ m \rightarrow \infty $  for macroscopic systems.}  
 
 \begin{equation}
 \lim_{\substack{\hbar \to 0 \\ m \to \infty}} Q_G=\lim_{\substack{ \hbar \to 0 \\ m \to \infty}} \left[  \frac{\hbar^2}{4m\beta^2}(1-\frac{x^2}{\beta^2}) \right] \approx 0 \label{limit}
 \end{equation}
 
\noindent  This finally suggests that the center of mass of a macroscopic Bohmian system in interaction with the environment--a macroscopic system represented by an ES-EWF--will move according to Newton's second law of dynamics \eqref{classicalNewton}.






\section{Conclusions}
The paper shows that Bohm's theory reduces to Newton's theory in the macroscopic classical regime. When a Bohmian system interacts with an external environment, effective wave functions are dynamically selected by the the environment (ES-EWFs). The derivation of the classical Newtonian dynamics follows from two main considerations: (i) the ES-EWFs can be identified with the pointer states of standard decoherence and, in the macroscopic regime, they are Gaussian states; (ii) the quantum potential of a Gaussian state is negligible when the system is massive and $\hbar \rightarrow 0$. Both of these conditions apply when the system is macroscopic. Therefore, a macroscopic Bohmian system in interaction with the environment will follow an approximately Newtonian dynamics.       

\section*{Acknowledgments}    
I wish to thank Mario Hubert for his helpful feedbacks and comments on previous drafts of this paper. I also want to thank Valia Allori, Guido Bacciagaluppi, Andrea Oldofredi, Patricia Palacios and Antonio Vassallo with whom I discussed the topic presented in this work on many occasions over the years. This work has been supported by the \emph{Funda\c c\~ao para a Ci\^encia e a Tecnologia} through the fellowship FCT Junior Researcher, hosted by the Centre of Philosophy of the University of Lisbon.


\clearpage

\begin{thebibliography}{99}
\bibitem{} L. E. Ballentine (2015), \emph{Quantum Mechanics: A Modern Development}, 2nd edition, World Scientific Publishing, Singapore. 
\bibitem{} D. Bohm (1952a), A suggested interpretation of the quantum theory in terms of ``hidden'' variables I, \emph{Physical Review}, 85(2), 166-179.
\bibitem{} D. Bohm (1952b), A suggested interpretation of the quantum theory in terms of ``hidden'' variables II, \emph{Physical Review}, 85(2), 180-193.
\bibitem{} D. Bohm \& B. Hiley (1987), An ontological basis for the quantum theory, part I: non-relativistic particle systems, \emph{Physics Reports}, 144(6): 323-348.
\bibitem{} D. Bohm \& B. Hiley (1993), \emph{The Undivided Universe: An Ontological Interpretation of Quantum Theory}, Routledge. 
\bibitem{} L. Diosi \& C. Kiefer (2000), Robustness and diffusion of pointer states, \emph{Physical Review Letters} 85, 3552. 
\bibitem{} A. Drezet,(2021)  Justifying Born's Rule Using Deterministic Chaos, Decoherence, and the de Broglie-Bohm Quantum Theory. \emph{Entropy}, vol. 23: 1371. 
\bibitem{} D. D\"urr, S. Goldstein \& N. Zanghi (1992), Quantum equilibrium and the origin of absolute uncertainty, \emph{Journal of Statistical Mechanics}.  
\bibitem{} D. D\"urr \& S. Teufel (2009), \emph{Bohmian Mechanics: The Physics and Mathematics of Quantum Theory}, Springer, Berlin. 
\bibitem{} A. Einstein (1953), Elementare Uberlegungen zur Interpretation der Grundlagen der Quanten-Mechanik, in Scientific Papers Presented to Max Born on his Retirement from the Tait Chair of Natural Philosophy in the
University of Edinburgh (Edinburgh, Oliver and Boyd), pp. 33-40.
\bibitem{} P. R. Holland (1993), The Quantum Theory of Motion: An Account of the de Broglie-Bohm Causal Interpretation of quantum mechanics, Cambridge University Press, Cambridge.
\bibitem{} E. Joos, H.  Zeh, C. Kiefer, D. Giulini, J. Kupsch, \& I. Stamatescu (2013). \emph{Decoherence and the Appearance of a Classical World in Quantum Theory}, Springer.
\bibitem{} W. Myrvold (2003), On some early objections to Bohm's theory, \emph{International Studies in the Philosophy of Science}, 17(1). 
\bibitem{} T. Norsen (2018), On the explanation of Born-rule statistics in the de Broglie-Bohm pilot-wave theory, \emph{Entropy}, 20(6): 422.
\bibitem{} D. Romano (2016), Bohmian classical limit in bounded regions, in L. Felline, A. Ledda, F. Paoli and E. Rossanese (eds.): \emph{New Directions in Logic and the Philosophy of Science}, SILFS series, vol. 3, College Publications, London, 2016.
\bibitem{} D. Romano (2021), Multi-field and Bohm's theory, \emph{Synthese}, 198(11): 10587-10609.
\bibitem{} D. Romano (2022), The unreasonable effectiveness of decoherence, in V. Allori (ed.): \emph{Quantum Mechanics and Fundamentality: Naturalizing Quantum Theory between Scientific Realism and Ontological Indeterminacy}, vol. 460: 3-18, \emph{Synthese Library}, Springer. 
\bibitem{} C. Rovelli (2022), Preparation in Bohmian mechanics, \emph{Foundations of Physics}, 52(3): 59. 
\bibitem{} M. A. Schlosshauer (2007), \emph{Decoherence and the quantum to classical transition}, Springer, Berlin.
\bibitem{} M. A. Schlosshauer (2019), Quantum decoherence, \emph{Physics Reports}, 831: 1-57.  
\bibitem{} L. Sorgel \& K. Hornberger (2015), Unraveling quantum brownian motion: pointer states and their classical trajectories, \emph{Physical Review A}, 92, 062112.
\bibitem{}  A. Valentini (1991), Signal-locality, uncertainty, and the subquantum H-theorem I, \emph{Physics Letters A}, 156(1-2): 5-11.
\bibitem{} W. H. Zurek (1981), Pointer basis of quantum apparatus: Into what mixture does the wave packet collapse?, \emph{Physical Review D},24: 516-1525.
[\bibitem{}  W. H. Zurek (1982), Environment-induced superselection rules, \emph{Physical Review D} : 26: 862-1880.
\bibitem{} W. H. Zurek (2002), Decoherence and the transition from quantum to classical--revisited, \emph{Los Alamos Science}, n. 27. Updated version of \emph{Physics Today} (1991), n. 44: 36-44. 
\bibitem{} W. H. Zurek, S. Habib \& J. P. Paz (1993), Coherent states via decoherence, \emph{Physical Review Letters}, 70(9), 1187-1190.


\end{thebibliography}
\end{document}